\documentclass{webofc}
\usepackage[varg]{txfonts}   
\usepackage{graphicx} 

\title{HEPScore: A new CPU benchmark for the WLCG}

\author{
\firstname{Domenico} \lastname{Giordano} \inst{1} \fnsep\thanks{\email{Domenico.Giordano@cern.ch}}
\and
\firstname{Jean-Michel} \lastname{Barbet} \inst{2}
\and
\firstname{Tommaso} \lastname{Boccali}\inst{3}
\and
\firstname{Gonzalo Menéndez} \lastname{Borge} \inst{1}
\and
\firstname{Christopher} \lastname{Hollowell} \inst{4} 
\and
\firstname{Vincenzo} \lastname{Innocente} \inst{1}
\and
\firstname{Walter} \lastname{Lampl} \inst{5}
\and
\firstname{Michele}  \lastname{Michelotto} \inst{6} 
\and
\firstname{Helge} \lastname{Meinhard} \inst{1}
\and
\firstname{Ladislav} \lastname{Ondris} \inst{1}
\and
\firstname{Andrea} \lastname{Sciab\`{a}} \inst{1}
\and
\firstname{Matthias J.} \lastname{Schnepf} \inst{7}
\and
\firstname{Randall J.} \lastname{Sobie} \inst{8}
\and
\firstname{David} \lastname{Southwick} \inst{1}
\and
\firstname{Tristan S.} \lastname{Sullivan} \inst{8}
\and
\firstname{Andrea} \lastname{Valassi} \inst{1}
\and
\firstname{Sandro} \lastname{Wenzel} \inst{1}
\and
\firstname{John L.} \lastname{Willis} \inst{9}
\and
\firstname{Xiaofei} \lastname{Yan} \inst{10}
}
\institute{
CERN, Geneva, Switzerland
\and
Subatech UMR 6457, CNRS-IN2P3, IMT Atlantique, Université de Nantes,  Nantes, France
\and
INFN Sezione di Pisa, L.go Pontecorvo 3, 56127 Pisa (ITALY), and ICSC-National Center for HPC, Big Data and Quantum Computing, Italy
\and
Brookhaven National Laboratory, United States
\and
University of Arizona, United States
\and
INFN, Istituto Nazionale di Fisica Nucleare, Padua, Italy
\and
Karlsruhe Institute of Technology, Karlsruhe, Germany
\and
Institute of Particle Physics and University of Victoria, Victoria, British Columbia, Canada
\and
California Institute of Technology, Pasadena, California, United States
\and
Institute of High Energy Physics, Beijing, China
}

\abstract{%
HEPScore is a new CPU benchmark created to replace the HEP-SPEC06 benchmark that is currently used by the WLCG for procurement,  computing resource pledges, usage accounting and performance studies.
The development of the new benchmark, based on HEP applications or workloads,  has involved many contributions from software developers, data analysts, experts of the experiments, representatives of several WLCG computing centres and WLCG site managers.
In this contribution, we review the selection of workloads and the validation of the new HEPScore benchmark.
}

\begin{document}
\maketitle

\section{Introduction}

Computing in particle physics has evolved to a highly distributed model where each country provides local facilities that are integrated into a global infrastructure, called the Worldwide LHC Computing Grid (WLCG)~\cite{WLCG}.
The WLCG coordinates the computing and networking resources on behalf of the experiments at the Large Hadron Collider at CERN.

The collaborative nature of our field has resulted in agreements for the sharing of costs for all aspects of our experiments, including computing, either through direct financial payments or the provision of in-kind equipment or services.
Each country is requested to contribute resources based on the size of their research community and financial oversight boards find consensus on the appropriate cost sharing.

It is difficult to put a cost estimate on a computing facility as each country has its own way of acquiring and operating their resources.
Rather than develop a model based on cost of the facility, hardware and personnel, it was decided to use the CPU-power delivered by a site to compare the CPU resources provided by each country.
The delivered CPU-power is defined to be the number of seconds used by the applications multiplied by a benchmark that reflects the performance of the servers.
After many years of operations, most sites have many types of servers with differing levels of performance.
As a result, the sites report a benchmark that is average of the individual benchmarks of the different servers weighted by number of servers of each type (we refer to this number as the {\it site benchmark}).

The resources used at each site are tracked and stored in a WLCG accounting database.
The database stores the site benchmark and the number of CPU-seconds used by each experiment at a site.
These numbers are used to calculate an integrated number that estimates the resources delivered by the site.
These numbers are published by the WLCG accounting team on a monthly basis and provided to the funding agencies.

\section{Motivation for a new CPU benchmark}
In 2009, the WLCG agreed to use HEP-SPEC06 (HS06) as its CPU benchmark~\cite{HEPIX_Bmk_WG}. 
HS06 is based on the industry-standard SPEC CPU 2006 benchmark~\cite{SPEC2006}.
At that time, the HEP applications shared several commonalities with a number of workloads within the SPEC CPU 2006 suite and those workloads were selected to be included in HS06~\cite{michelotto_comparison_2010, Giordano:2020qhu}.
The individual HS06 workloads were single-threaded and single-process applications, compiled in 32-bit mode, and requiring a minimum of 1 GB of memory per process. 

\begin{figure}[h]
\centering
\includegraphics[width=12cm,clip]{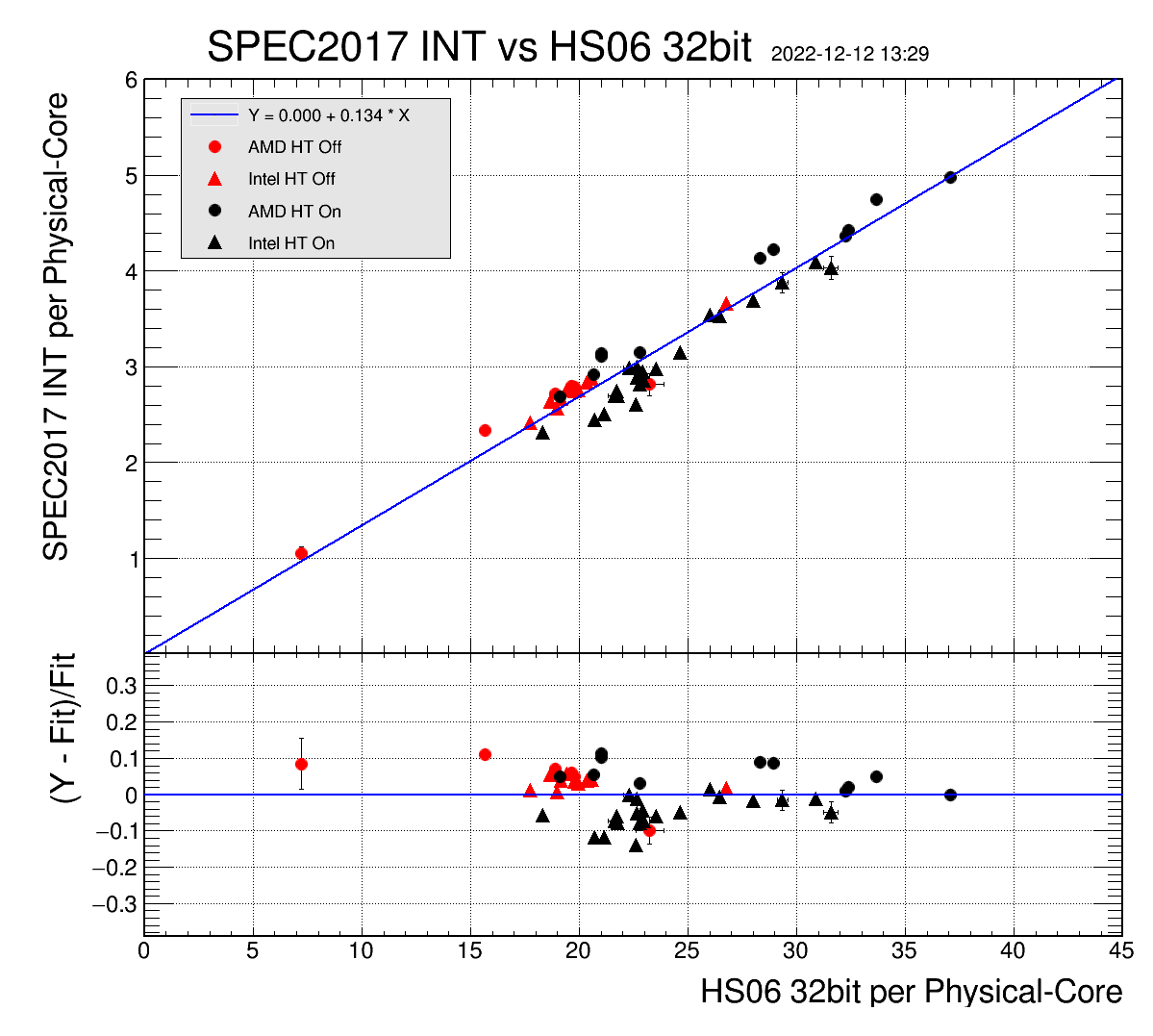}
\caption{A comparison of the HEP-SPEC06 benchmark with the results of the SPEC CPU 2017 benchmark on resources provided to the HEPiX Benchmark Working Group.
The circles are measurements on AMD processors and triangles are on Intel processors. 
The red (black) points have hyper-threading off (on).
The blue line is a linear fit to the data constrained to the origin (0,0).
The lower plot shows the fractional difference in the vertical dimension of the data point from the fit.
Note that the benchmarks are normalized to the number of physical cores of the server 
independent of whether hyper-threading is enabled.
}
\label{fig:hs06}       
\end{figure}

The support of the SPEC CPU 2006 benchmark has ended and the Working Group considered a number of options for replacement for HS06.
A newer release of the SPEC benchmark, SPEC CPU 2017~\cite{SPEC2017}, was also studied but found to be highly correlated with the HEP-SPEC06 benchmark (see fig.~\ref{fig:hs06}).  
The SPEC CPU 2017 benchmark would require each site to purchase a new licence (like the SPEC CPU 2006 benchmark) whereas the WLCG community was in favour an open-source benchmark.
As a result, the SPEC CPU 2017 was rejected as a replacement for HS06.

HS06 has met the WLCG requirements in a world that has progressively evolved from CPUs with few cores to multi-core CPUs. 
During this period, experiments were starting to observe that newer versions of their applications scaled less well with HS06 (for example, see ref.~\cite{Charpentier}). 
The proposal for building a new benchmark using HEP applications was first presented at the 2017 WLCG workshop~\cite{wlcg:manchester}.
The HEPiX Benchmarking Working Group\footnote{HEPiX Benchmarking Working Group: D~Giordano (co-chair), M~Michelotto (co-chair), L~Atzori, JM~Barbet, GM~Borge, C~Hollowell, L~Ondris, A~Sciaba, E~Simili, RJ~Sobie, D~Southwick, T~Sullivan, N~Szczepanek, A~Valassi} was asked to study the feasibility of a new benchmark based on HEP workloads.
This led to the creation of the HEP Benchmarks project and the development of a software framework to benchmark the performance of computing servers using HEP applications~\cite{valassi}.
The Working Group developed the {\it HEP Benchmark Suite} that provides a simple way to run containerized HEP applications and other benchmarks, and record the results in an Elasticsearch database at CERN~\cite{CSBS-Benchmark}.

Some of the HEP applications that use the largest amount of compute power are the Monte Carlo generation of collision events,  the simulation of the detector response to the simulated particles, the conversion of the simulated energy deposition by the particles in the detector elements (digitization) and reconstruction of the detector signals into particles and momenta (the reconstruction code is used for both simulated and real data).  
We refer to these HEP applications as {\it workloads}.
Analysis applications are very specific to the physics, and often I/O intensive, and are considered too unreliable to use as a measure of the performance of a CPU.

The creation of a new benchmark based on HEP workloads (called {\it HEPScore}) benefits from the consensus of the WLCG community.
As a result, the  WLCG Management Board established the HEPScore Benchmark Deployment Task Force\footnote{WLCG HEPscore Deployment Task Force: D~Giordano (co-chair), RJ~Sobie (co-chair), 
J~Andreeva, G~Andronico, GM~Arevalo, T~Boccali, GM~Borge, C~Bozzi, S~Campana, I~Collier, 
A~Di~Girolamo, M~Jouvin, W~Lampl, JR~Letts, Z~Marshall, H~Meinhard, AM~Melo, B~Panzer-Steindel, 
S~Piano, D~Piparo, F~Qi, MJ~Schnepf, O~Smirnova, J~Templon, A~Valassi, JL~Willis, T~Wong} whose role was to review the requirements for the HEPScore benchmark and to help select the HEP workloads that are to be used in the HEPScore benchmark.
The Task Force was also asked to propose a transition plan for the migration from the HS06 benchmark to the new HEPScore benchmark.

In September 2022, a 2-day workshop at CERN devoted to the HEPScore benchmark brought together members
of the Working Group, Task Force, computing coordinators and software developers of the experiments and site representatives to discuss the composition of the new benchmark~\cite{hepscore:workshop}.
A presentation at the 2022 ACAT conference provided an interim report on the status of the HEPScore benchmark~\cite{Giordano:acat2022}.
The proposal for the first version of HEPScore was presented at the 2022 WLCG Workshop~\cite{wlcg:sobie} where a number of recommendations were proposed (discussed in the next section).
A meeting of the WLCG Management Board reviewed the status of the new benchmark in December 2022 and set a milestone of April 2023 for the release of HEPScore.

Concurrent to the development of the HEPScore benchmark, the issue of power consumption and environmental impact of computing resources has become a topical area of interest~\cite{wlcg:campana}.
A study, using a beta version of HEPScore, evaluated the processing capabilities and power consumption of x86 and ARM processors, and showed that ARM processors use less power while still being highly performant for HEP workloads~\cite{acat:simili}.
As a result, there was strong community interest in a benchmark for both x86 and ARM processors, and a concerted effort by the experiments resulted in all workloads being ready for both processor types for the April 2023 milestone.

\section{HEP Workloads}

In 2021 and 2022, workloads were provided to the Working Group by all four large LHC experiments (ALICE, ATLAS, CMS and LHCb) and other experiments (Belle~II, JUNO and the International Gravitational Wave Network).

\begin{figure}[h]
\centering
\includegraphics[width=4.25cm,clip]{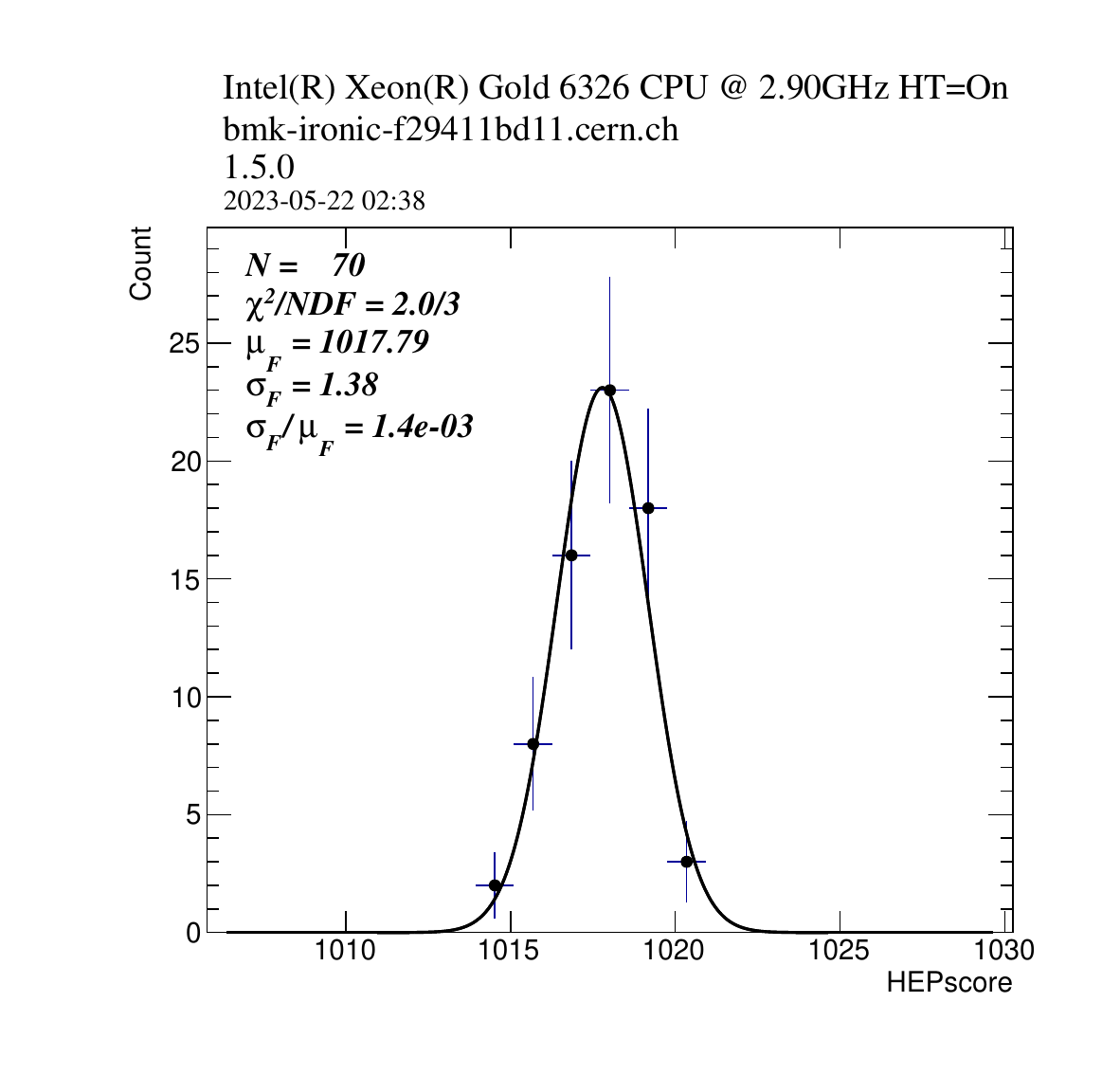}
\includegraphics[width=4.25cm,clip]{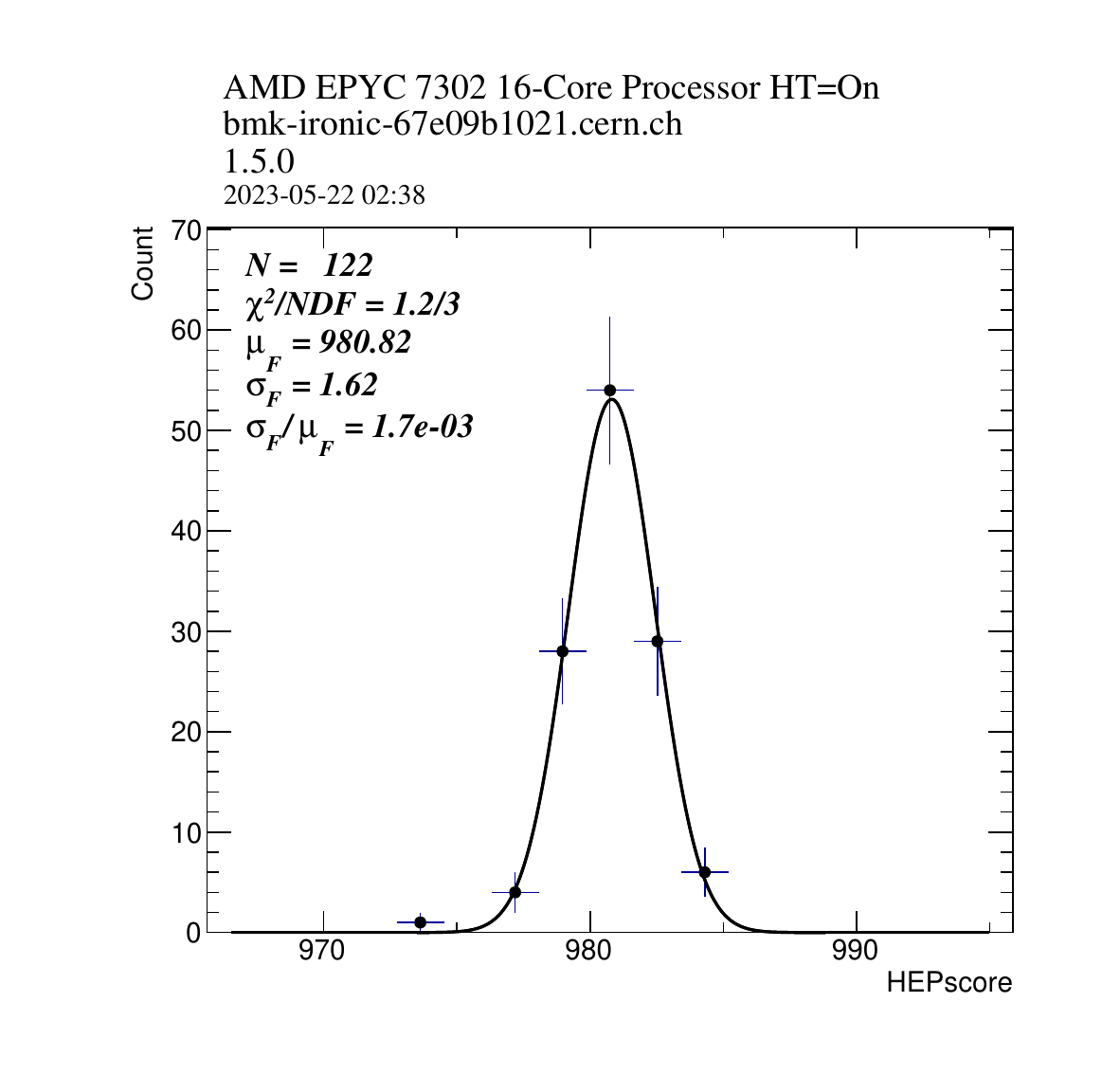}
\includegraphics[width=4.25cm,clip]{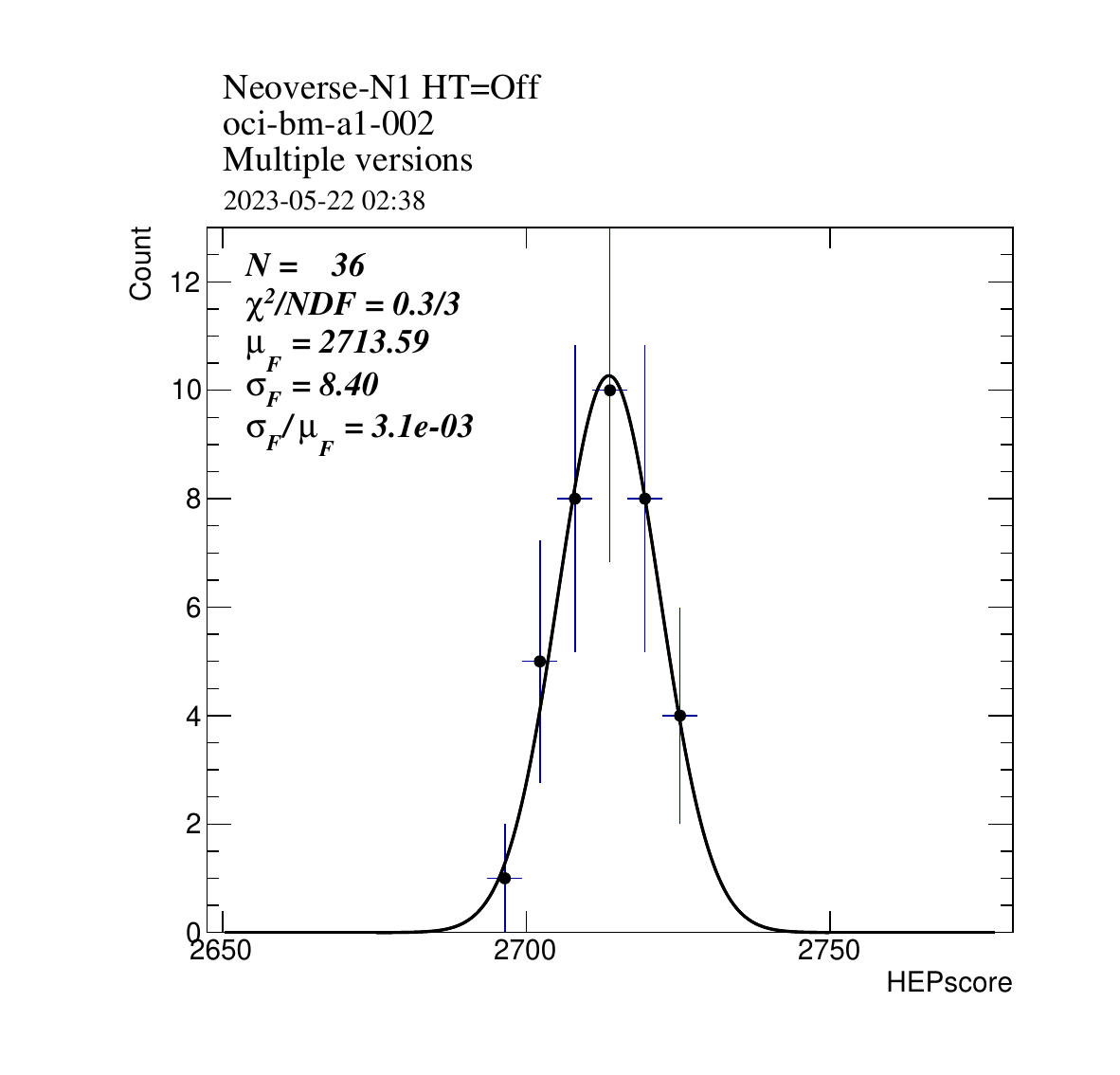}
\caption{Histograms of the HEPScore23 benchmark on an Intel, AMD and ARM processor.  The fits to the histogram use a Gaussian distribution.}
\label{fig:HS32-fits}       
\end{figure}

Each workload is encapsulated into a container with the software and input data needed to run the application.
The software of the experiment is stored in the CVMFS file system~\cite{cvmfs} and then made accessible via a local folder inside the container. 
The set of containers of workloads is stored in a Gitlab repository at CERN.
Each container includes a configuration file with a parameter for the number of events that allows one to adjust the duration of the execution (more details can be found in ref.~\cite{CSBS-Benchmark}).
Each workload is run three times and the geometric mean is taken as the benchmark 
(typically in units of events per second);  this is identical to the method used for each workload component in HS06.

\begin{figure}[h]
\centering
\includegraphics[width=10cm,clip]{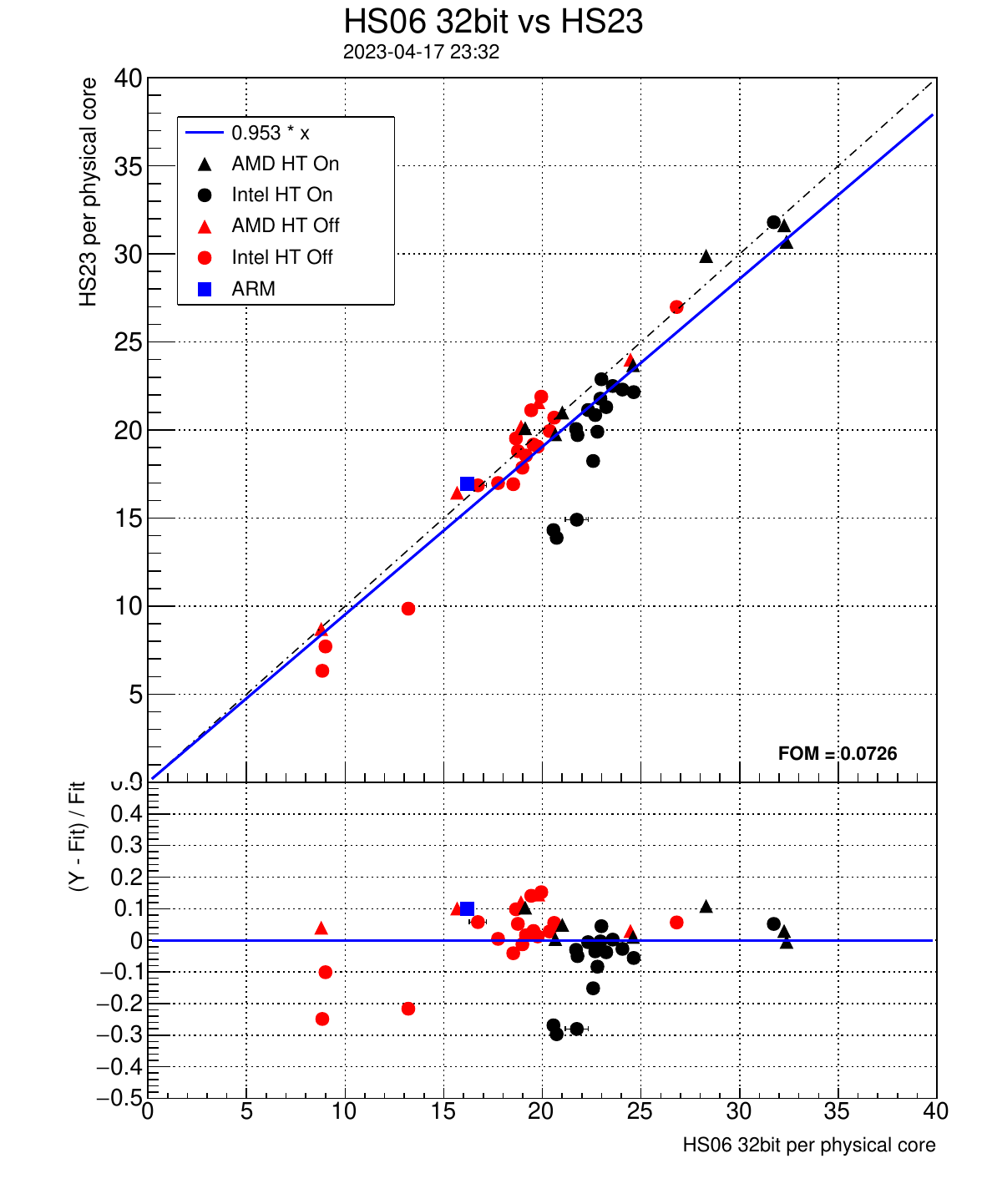}
\caption{The blue line is a linear fit to the data points constrained to the origin and the dashed line has unity slope. 
The circles are measurements on Intel processors, the triangle points are measurements on AMD processors and the box point is an ARM processors measurement.
The points in red (black) where taken with hyper-threading off (on).
The measurement on the single ARM processor is with hyper-threading off (ARM processors are not designed to operate with hyper-threading).}
\label{fig:HS32-vs-HS06}       
\end{figure}

Each workload was validated on a set of dedicated servers at CERN to check the reliability and reproducibility, and in all cases, the results were found to be consistent at a level better than 1\%.
Once validated, the workloads were run on a large and diverse set of server systems provided by many WLCG sites.
The benchmarks of the individual workloads from the server systems were compared to determine the correlations between applications.
The time to run the each workload and the studies of the correlations provided valuable input that was used to help select the workloads that are included in HEPScore.

At the Benchmark Workshop, the criteria for selecting workload candidates for HEPScore were discussed.
The conclusion was that HEPScore should be representative of the shares of computing usage of the experiments, it should run in a timely manner (3-6 hours) and provide complementary workloads (e.g. avoid the selection of highly correlated workloads).
It is important that the workloads from the experiments be stable for a reasonable period (e.g. 2-3 years), and ideally, valid for at one LHC run period.

Seven workloads were selected to be part of the HEPScore23 benchmark (the benchmark is called HEPScore23 to indicate the year the benchmark was created).
HEPScore23 includes two workloads each from CMS (reconstruction and generation-simulation) and ATLAS (Sherpa-generation, reconstruction), and workloads from ALICE (digitization-reconstruction), Belle~II (generation-simulation-reconstruction) and LHCb (simulation).  
The workloads were chosen to cover diverse types of applications, to be complementary in terms of scaling across server types, to run in an acceptable time, and to represent complex event topologies.
The time to run each workload ranges from 300 to 900 seconds on the reference server at CERN (Intel Xeon Gold 6326 CPU @ 2.90GHz).
HEPScore23 follows the methods used for HEP-SPEC06 of running each workload three times and taking the geometric mean of the three measurements.
The total time to run the HEPScore23 is approximately 3.5 hours on the reference server.

The workloads in HEPScore23 are equally weighted.  
The Working Group studied different weighting schemes but found little difference.
The choice of equal weighting of the workloads gives additional impact to ATLAS and CMS as they each provide two workloads.

As part of the validation process, the HEPScore23 benchmark was measured on Intel, AMD and ARM servers at CERN.  
The results, shown in fig.~\ref{fig:HS32-fits}, demonstrate the reproducibility of the HEPScore23 benchmark on the different architectures to better than 1\%.
The HEPScore23 benchmark was run on a wider set of servers with both hyper-threading on and off.
In fig.~\ref{fig:HS32-vs-HS06}, the HEPScore23 benchmark is plotted against the HEP-SPEC06 (32-bit version). 

\begin{figure}[h]
\centering
\includegraphics[width=13cm,clip]{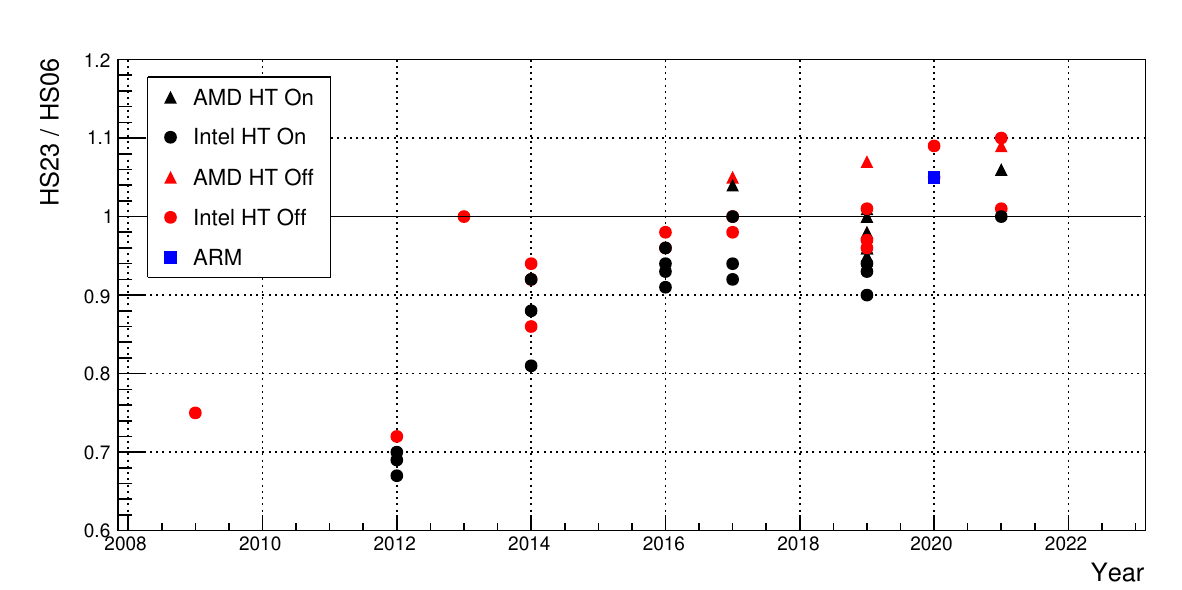}
\caption{Plot of the ratio of the HEPScore23 to HEP-SPEC06 benchmarks as a function of the year of the release of the server model.
The circles are measurements on Intel processors, the triangle points are measurements on AMD processors and the box point are 
on ARM processors.
The points in red (black) were taken with hyper-threading off (on).
The measurement on the single ARM processor is with hyper-threading off (ARM processors are not designed to operate with hyper-threading).}
\label{fig:year}       
\end{figure}

In fig.~\ref{fig:year}, we show the ratio of the HEPScore23 to HEP-SPEC06 as a function of the year in which the processor was introduced on the market (the colours of the points are identical to those used in fig.~\ref{fig:HS32-vs-HS06}).
The measurement of HEPScore23 was normalized to the value HEP-SPEC06 on the reference machine (the data point for the reference machine is one of the points in 2021).
It is observed that the ratio of HEPScore23 to HEP-SPEC06 is less than unity for older machines and increases for newer servers.

\section{Summary}

The HEPScore23 benchmark was released for use by WLCG sites in April 2023. 
HEPScore23 is normalized to HEP-SPEC06 as measured on the reference machine to facilitate an easy transition for the sites and the WLCG accounting group.
Sites are being asked to use HEPScore23 (or both benchmarks) to evaluate newly procured hardware.
Existing hardware will not need to be benchmarked with HEPScore23 and sites can continue to use their current benchmarks based on HEP-SPEC06.
The WLCG Management Board will review the situation and decide whether to make HEPScore23 the required benchmark for future years.

\end{document}